\documentstyle[preprint,aps,epsf]{revtex}
\begin{document}

\draft
\title{
Calculation of the average Green's function of electrons in a stochastic medium
via higher-dimensional bosonization}

\author{Peter Kopietz}
\address{
Institut f\H{u}r Theoretische Physik der Universit\H{a}t G\H{o}ttingen,\\
Bunsenstr.9, D-37073 G\H{o}ttingen, Germany}
\date{August 9, 1995}
\maketitle
\begin{abstract}

The disorder averaged single-particle Green's function of
electrons subject to a time-dependent 
random potential with long-range spatial correlations is calculated by means of bosonization
in arbitrary dimensions.
For static disorder our method is equivalent with
conventional perturbation theory based on the lowest order Born approximation.
For dynamic disorder, however,
we obtain a new non-perturbative expression for the average Green's function. 
Bosonization also provides a solid microscopic basis for the description of the quantum dynamics 
of an interacting many-body system  via an effective stochastic model
with Gaussian probability distribution.

\end{abstract}

\narrowtext

Over the past 20 years
the motion of a quantum particle in a time-dependent random potential 
has been the subject of many works\cite{Ovchinnikov74,Girvin79}. 
Most authors have focused on the case of a single isolated electron,
where numerical- as well as non-perturbative analytical methods are 
available\cite{Ovchinnikov74}.
The equally important problem of 
calculating the average single-particle Green's function 
in the presence of a filled Fermi sea has 
not received much attention\cite{Girvin79}. 
The purpose of the present paper is to show that the 
functional integral formulation of 
bosonization\cite{Fogedby76,Frohlich95,Kopietz94a,Kopietz94b,Kopietz95} 
offers a new non-perturbative approach to this problem
in arbitrary dimensions $d$.

The bosonization approach in $d > 1$ has recently been 
re-discovered by Haldane\cite{Haldane92}, who 
generalized earlier ideas due to Luther\cite{Luther79}. 
For a detailed description of the geometric construction leading to
higher-dimensional bosonization we would like to refer the reader to 
Ref.\cite{Haldane92}. 
Here we briefly recall the basic features of Haldane's construction.
The first step is the subdivision of the
Fermi surface into disjoint patches $\tilde{K}^{\alpha}_{\Lambda}$
and the introduction of 
a collection of
local coordinate systems with origins on the Fermi surface 
(this is called an {\it{atlas}}\cite{Nash83}).
In $d$ dimensions each patch covers an area $\Lambda^{d-1}$ of the Fermi surface, where
the cutoff $\Lambda$ should be chosen such that 
within a given patch the curvature of the Fermi surface can be locally neglected.
For spherical Fermi surfaces, this means that $\Lambda$ should be chosen small
compared with the Fermi wave-vector $k_{F}$.
Each patch is then extended into a $d$-dimensional box $K^{\alpha}_{\Lambda , \lambda}$
with radial height $\lambda$ and  volume $\Lambda^{d-1} \lambda$. 
One  proceeds by defining local density operators associated with the boxes,
$\hat{\rho}^{\alpha}_{\bf{q}} = \sum_{\bf{k}} \Theta^{\alpha} ( {\bf{k}} )
\hat{c}^{\dagger}_{\bf{k}} \hat{c}_{ {\bf{k}} + {\bf{q}} }$, where
$\hat{c}_{ \bf{k}} $ is the annihilation operator of an electron with wave-vector
${\bf{k}}$ (for simplicity we consider spinless electrons in this work), and the
cutoff function $\Theta^{\alpha} ( {\bf{k}} )$  is 
unity for ${\bf{k}} \in K^{\alpha}_{\Lambda , \lambda} $ and vanishes otherwise.
In the functional bosonization 
approach\cite{Fogedby76,Frohlich95,Kopietz94a,Kopietz94b,Kopietz95} 
the calculation of the interacting Green's function 
is then mapped via a Hubbard-Stratonowich transformation
onto the problem of calculating the  {\it{average}} Green's function of an effective
non-interacting system in a dynamic random potential
$\phi^{\alpha}$,
which couples to the local densities 
$\hat{\rho}^{\alpha}$.
Because in the high-density limit the probability
distribution of the $\phi^{\alpha}$-field is Gaussian, 
the calculation of the average Green's function reduces to the
calculation of a trivial Debye-Waller factor.
In the field theory literature\cite{Frohlich95} 
the auxiliary field $\phi^{\alpha}$ is called {\it{disorder field}}.
This terminology suggests that the functional integral formulation of 
bosonization can also be used to calculate the average Green's function
of non-interacting electrons in a dynamic random potential.
This connection between bosonization 
and disordered systems 
is obscured in the operator approach\cite{Houghton93,Castro94}, but it is
quite clear in the functional formulation. 
Indeed, in this work we shall show 
that the functional bosonization approach
developed in Refs.\cite{Kopietz94a,Kopietz94b,Kopietz95}
provides a useful method to calculate the average 
single-particle Green's function of electrons in a stochastic medium.

We consider non-interacting spinless electrons 
at high densities subject to 
an {\it{imaginary-time}}
random potential $U ( {\bf{r}} , \tau )$.
The imaginary-time Green's function 
${\cal{G}} ( {\bf{r}} , {\bf{r}}^{\prime} , \tau , \tau^{\prime} )$ 
is defined as usual,
 \begin{eqnarray}
 \left[ - \partial_{\tau} -  \frac{ ( - i \nabla_{\bf{r}}  )^2}{2m} + \mu  - U ( {\bf{r}} , \tau )
 \right] 
 {\cal{G}} ( {\bf{r}} , {\bf{r}}^{\prime} , \tau , \tau^{\prime} ) 
&  & 
 \nonumber 
 \\
 &  & \hspace{-30mm} =
 \delta ( {\bf{r}} - {\bf{r}}^{\prime} ) \delta^{\ast} ( \tau - \tau^{\prime} )
 \label{eq:Grrttdef}
 \; \; \; ,
 \end{eqnarray}
where $m$ is the mass of the electrons, $\mu$ is the average chemical potential, and
  $\delta^{\ast} ( \tau ) = \frac{1}{\beta} \sum_{n} e^{ - i \tilde{\omega}_{n} \tau }$.
Here $\beta$ is the inverse temperature and $\tilde{\omega}_{n} = 2 \pi ( n + \frac{1}{2} ) / \beta$ 
are the fermionic Matsubara frequencies.
We assume that the random potential  has a Gaussian probability distribution with
zero average and general covariance 
 $\overline{ U ( {\bf{r}} , \tau )
 U ( {\bf{r}}^{\prime} , \tau^{\prime} ) }
  =  C  ( {\bf{r}} - {\bf{r}}^{\prime} , \tau - \tau^{\prime} )$,
where the overbar denotes averaging with respect to the probability distribution
of the random potential. 
In Fourier space we have then
 $\overline{ {U}_{q} {U}_{-q} } = \beta V C_{q}$,
where $V$ is the volume of the system, and the Fourier components are defined by
 \begin{equation}
 C_{q}  =  
 \int_{0}^{\beta} d \tau  \int d {\bf{r}}
 e^{ - i ( {\bf{q}} \cdot {\bf{r}} - \omega_{m} \tau )} C ( {\bf{r}} , \tau ) 
 \; \; \; ,
 \end{equation}
and analogously for $U_{q}$.
Here $q = [ {\bf{q}} , i \omega_{m} ]$, and
$\omega_{m} =  2 \pi m / \beta$ are the bosonic Matsubara frequencies.

We are interested in the average Green's function
 $G ( {\bf{r}} - {\bf{r}}^{\prime} , \tau - \tau^{\prime} )
 = \overline{ 
 {\cal{G}} ( {\bf{r}} , {\bf{r}}^{\prime} , \tau , \tau^{\prime} ) 
 }$.
Within our bosonization approach 
 $G ( {\bf{r}} - {\bf{r}}^{\prime} , \tau - \tau^{\prime} )$
is calculated in the most direct way: First we obtain the exact
Green's function 
 ${\cal{G}} ( {\bf{r}} , {\bf{r}}^{\prime} , \tau , \tau^{\prime} ) $ for
a given realization of the random potential, and then this expression is averaged.
As explained in detail in Ref.\cite{Kopietz95}, the first step can only
be carried out if there exists a cutoff $q_{c} \ll k_{F}$ such that 
$U_{q}$ becomes negligibly small for 
$|{\bf{q}}| 
 \raisebox{-0.5ex}{$\; \stackrel{>}{\sim} \;$} 
 q_{c}$.
But this means that we have to assume that for
wave-vectors $ | {\bf{q}} | \raisebox{-0.5ex}{$\; \stackrel{>}{\sim} \;$} q_{c}$ 
the Fourier coefficients
$C_{q}$ of the covariance function can be neglected. 
In other words, we can only study random potentials
with sufficiently long-range spatial correlations, such that
$q_{c} \ll k_{F}$.
Evidently the most popular model of $\delta$-function correlated disorder 
cannot be treated within our bosonization approach.
However, in view of the fact that
a random potential with a finite correlation range $q_{c}^{-1 }$ is expected to
lead for distances $| {\bf{r}} | \gg q_{c}^{-1} $
to qualitatively identical results for single-particle
properties as a $\delta$-function
correlated random potential, this restriction seems not to be very serious.
By choosing the above mentioned box cutoffs 
such that $q_{c} \ll \Lambda , \lambda \ll k_{F}$, we obtain
cutoff-independent results for physical correlation functions
at distances $ | {\bf{r}} | \gg q_{c}^{-1}$.

Due to the subdivision of the Fermi surface into small patches  
it is possible to linearize the energy dispersion
locally within a given patch,
$[  \frac{ {\bf{k}}^2}{2 m} - \mu ]_{ {\bf{k}} = {\bf{k}}^{\alpha} + {\bf{q}} }
 \approx {\bf{v}}^{\alpha} \cdot {\bf{q}}$,
where ${\bf{k}}^{\alpha}$ is the center of patch
$\tilde{K}^{\alpha}_{\Lambda}$, and ${\bf{v}}^{\alpha}$ is the local Fermi velocity.
We then replace 
Eq.(\ref{eq:Grrttdef}) by  
a {\it{first order}} partial differential equation for
the patch Green's function
 ${\cal{G}}^{\alpha} ( {\bf{r}} , {\bf{r}}^{\prime} , \tau , \tau^{\prime} ) $, 
 \begin{eqnarray}
 \left[ - \partial_{\tau} + i {\bf{v}}^{\alpha} \cdot \nabla_{\bf{r}}    - U ( {\bf{r}} , \tau )
 \right] 
 {\cal{G}}^{\alpha} ( {\bf{r}} , {\bf{r}}^{\prime} , \tau , \tau^{\prime} ) 
 & = &
 \nonumber
 \\
 & & \hspace{-30mm}
 \delta ( {\bf{r}} - {\bf{r}}^{\prime} ) \delta^{\ast} ( \tau - \tau^{\prime} )
 \label{eq:Grrttpatch}
 \; \; \; .
 \end{eqnarray}
Because this  differential equation is first order and linear,
it can be solved exactly by means of a trivial generalization of
Schwinger's ansatz\cite{Schwinger62}.  
The result can be written as
 \begin{eqnarray}
 {\cal{G}}^{\alpha} ( {\bf{r}} , {\bf{r}}^{\prime} , \tau , \tau^{\prime} ) 
 & = &
 G_{0}^{\alpha} ( {\bf{r}} - {\bf{r}}^{\prime} , \tau - \tau^{\prime} )
 \nonumber
 \\
 &  & \hspace{-20mm} \times
 \exp \left[  \frac{1}{\beta V} \sum_{q} U_{q} \frac{ e^{i [ {\bf{q}} \cdot {\bf{r}} - \omega_{m} \tau ]}
 -  e^{ i [ {\bf{q}} \cdot {\bf{r}}^{\prime} - \omega_{m} \tau^{\prime} ]}}
 { i \omega_{m} - {\bf{v}}^{\alpha} \cdot {\bf{q}} } \right]
 \; \; \; ,
 \label{eq:Grtaudis}
 \end{eqnarray}
with
 \begin{equation}
 G^{\alpha}_{0} ( {\bf{r}}  , \tau  )
 = 
 \delta^{(d-1)} ( {\vec{r}}^{\alpha}_{\bot}  )
 \left( \frac{ - i}{2 \pi} \right)
 \frac{1}
 { 
 r^{\alpha}  
 + i | {\bf{v}}^{\alpha} |  \tau }
 \; \; \; ,
 \label{eq:Gpatchreal1}
 \end{equation}
where
 $\delta^{(d-1)} ( {\vec{r}}^{\alpha}_{\bot}  )$
is a Dirac $\delta$-function of the $d-1$ components
$\vec{r}^{\alpha}_{\bot}$ of ${\bf{r}}$
that are orthogonal to ${\bf{v}}^{\alpha}$.
The Gaussian average of Eq.(\ref{eq:Grtaudis}) yields the
usual Debye-Waller factor,
 \begin{eqnarray}
 \overline{ {\cal{G}}^{\alpha} ( {\bf{r}} , {\bf{r}}^{\prime} , \tau , \tau^{\prime} ) }
 & \equiv  &
 G^{\alpha} ( {\bf{r}} - {\bf{r}}^{\prime} , \tau - \tau^{\prime} )
 \nonumber
 \\
 & = &
 G_{0}^{\alpha} ( {\bf{r}} - {\bf{r}}^{\prime} , \tau - \tau^{\prime} )
  e^{  Q_{dis}^{\alpha} ( {\bf{r}} - {\bf{r}}^{\prime}, \tau - \tau^{\prime}) }
  \; \; \; ,
  \label{eq:Gavdis}
  \end{eqnarray}
with
 \begin{equation}
   Q_{dis}^{\alpha} ( {\bf{r}} , \tau  ) 
   =  
    -   \frac{1}{\beta V} \sum_{q} {C}_{q}
   \frac{ 1 - \cos ( {\bf{q}} \cdot {\bf{r}} - \omega_{m} \tau ) }
   { ( i \omega_{m} - {\bf{v}}^{\alpha} \cdot {\bf{q}} )^2 }
  \; \; \; .
  \label{eq:QdisDW}
  \end{equation}
The total disorder averaged Matsubara Green's function 
can then be written as 
 \begin{eqnarray}
 G ( {\bf{k}} , i \tilde{\omega}_{n} )  & = & \sum_{\alpha} \Theta^{\alpha} ( {\bf{k}} )
 \int d {\bf{r}} \int_{0}^{\beta} d \tau 
 e^{ - i [ ( {\bf{k}} - {\bf{k}}^{\alpha} ) \cdot  {\bf{r}}
 - \tilde{\omega}_{n}  \tau  ] }
 \nonumber
 \\
 & \times &
 {{G}}^{\alpha}_{0} ( {\bf{r}}  , \tau  )
 e^{ Q^{\alpha}_{dis} ( {\bf{r}} , \tau ) }
 \; \; \; .
 \label{eq:Gkresdis}
 \end{eqnarray}
Because $G^{\alpha}_{0} ( {\bf{r}} , \tau )$ is proportional to
$\delta^{d-1} ( \vec{r}^{\alpha}_{\bot} )$,
 we may replace
 $Q^{\alpha}_{dis} ( {\bf{r}} , \tau ) \rightarrow
 Q^{\alpha}_{dis} ( r^{\alpha} \hat{\bf{v}}^{\alpha} , \tau )$
in Eq.(\ref{eq:Gkresdis}), where
 $r^{\alpha} = \hat{\bf{v}}^{\alpha} \cdot {\bf{r}}$, with
 $\hat{\bf{v}}^{\alpha} = \frac{ {\bf{v}}^{\alpha} }{ | {\bf{v}}^{\alpha} | }$.
This completes the solution of the non-interacting problem.

Because disorder and interactions are treated on equal footing
in our bosonization approach,
it is easy to see that
interactions simply give rise to another {\it{additive}} contribution
$Q^{\alpha} ( {\bf{r}} , \tau )$ to the Debye-Waller factor. 
Hence, for disordered interacting electrons
Eq.(\ref{eq:Gkresdis}) should be modified by replacing  
$Q^{\alpha}_{dis} ( {\bf{r}} , \tau )$  by
 \begin{equation}
Q^{\alpha}_{tot} ( {\bf{r}} , \tau ) =
Q^{\alpha}_{dis} ( {\bf{r}} , \tau ) +
Q^{\alpha}_{int} ( {\bf{r}} , \tau ) 
\; \; \; ,
\label{eq:Qalphatot}
\end{equation}
where for a general (possibly retarded)
density-density interaction $f_{q}$ the contribution 
$Q^{\alpha}_{int} ( {\bf{r}} , \tau ) $ 
can be written as\cite{Kopietz94a,Kopietz94b,Kopietz95}
 \begin{equation}
 Q^{\alpha}_{int} ( {\bf{r}} , \tau ) =
    \frac{1}{\beta V}   \sum_{q} f^{RPA}_{q}
   \frac{ 1 - \cos ( {\bf{q}} \cdot {\bf{r}} - \omega_{m} \tau ) }
   { ( i \omega_{m} - {\bf{v}}^{\alpha} \cdot {\bf{q}} )^2 }
  \; \; \; .
  \label{eq:Qalphaint}
  \end{equation}
Here
 $f^{RPA}_{q} = f_{q} [ 1 + \Pi_{0} ( q ) f_{q} ]^{-1}$
is the random-phase approximation for the effective interaction, and
$\Pi_{0} ( q )$ is the non-interacting polarization.
In $d=1$ a result similar to Eqs.(\ref{eq:Qalphatot}) and (\ref{eq:Qalphaint})
has also been obtained by Hu and Das Sarma\cite{Hu93} by means
of a very different (and perhaps less elegant) method.
Evidently Eq.(\ref{eq:Qalphatot}) does not contain
interference terms describing weak localization effects. 
These are known to play an important role 
in the low-energy behavior of the average Green's function
of an interacting disordered Fermi system
in the metallic regime.
Note that diagrammatically weak localization is described 
in terms of infinite impurity ladders called
Cooperon-- and Diffuson propagators\cite{Lee85}, 
which satisfy the diffusion equation.
But the diffusive
motion involves large changes in the direction
of the particle due to  many successive (possibly small angle)
scatterings.
Such a motion cannot be correctly described
within the approximations inherent
in higher-dimensional bosonization
at the level of the Gaussian approximation\cite{Kopietz94a,Kopietz94b,Houghton93,Castro94}
(which amount (i) to the neglect of momentum-transfer between different
patches, and (ii) to the local linearization of 
the energy dispersion\cite{Kopietz95}),
because in this case the 
electron trajectories are approximated by
straight lines in the directions
of the local velocities ${\bf{v}}^{\alpha}$. 
Obviously the weak localization
effects must be contained in the corrections to the
straight line approximation for the electron trajectory.
Hence, the physics of weak localization will only emerge if
we can generalize our approach such that  
it describes changes in the direction of the
electron propagation due to successive scatterings.
This can be achieved {\it{either}}
by taking momentum-transfer between different
patches (so-called around-the-corner processes)
into account, {\it{or}} by considering
patches with a finite curvature 
(i.e. by retaining the quadratic terms in the expansion
of the energy dispersion at the Fermi 
surface)\cite{Giamarchi88}.
The equivalence of these procedures follows from the fact
that any patch with a finite  curvature can be subdivided
into a larger number of approximately flat sub-patches, such that
scattering from the original curved patch can also be described in terms
of scattering from the set of {\it{coupled}} but
flat sub-patches. The latter point of view leads in the
calculation of the Green's function for fixed
background field\cite{Kopietz95}
to a system of coupled first order differential
equations, which can always be transformed into a smaller system 
of differential equations with higher order derivatives.
At present it is not clear whether it is possible to 
include the above corrections into our formalism in
a simple approximate way, such that
we do not loose the physics of weak localization.
Obvious starting points might be to take
around-the-corner processes between neighboring patches into account,
or to retain second order spatial derivatives in
the partial differential equation (\ref{eq:Grrttpatch}).

Let us now show that
Eq.(\ref{eq:Gkresdis}) reduces for static disorder to the
usual perturbative result.
In this case only the $\omega_{m} = 0$- component of 
the covariance function $C_{q}$ is non-zero. 
For simplicity let us assume that $C_{q}$ has a separable form,
 $C_{q} = \delta_{\omega_{m} , 0 } \beta  \gamma_{0}
 e^{- \frac{ | {\bf{q}} |_{1} }{q_{c}} } $,
where $| {\bf{q}} |_{1} = \sum_{i=1}^{d} | q_{i} |$\cite{footnote1}.
Then we obtain from Eq.(\ref{eq:QdisDW}) for $ V \rightarrow \infty$
 \begin{equation}
 Q_{dis}^{\alpha} ( r^{\alpha} \hat{\bf{v}}^{\alpha} , \tau )
 \sim - \frac{ | r^{\alpha} | }{ 2 \ell^{\alpha} }
 \; \; \; , \; \; \; |r^{\alpha} q_{c} | \gg 1
 \; \; \; ,
 \label{eq:Qdisstatres}
 \end{equation}
where the inverse elastic mean free path is given by
 $\frac{1}{\ell^{\alpha}} = 
 \left( \frac{ q_{c}}{  {\pi}} \right)^{d-1}
 \frac{\gamma_{0}}{ | {\bf{v}}^{\alpha} |^2 }$.
We conclude that at large distances
 $G^{\alpha} ( {\bf{r}} , \tau ) = 
 G_{0}^{\alpha} ( {\bf{r}} , \tau )  e^{ - \frac{ |  r^{\alpha} | }{2 \ell^{\alpha} }}$.
In Fourier space this implies for $ | {\bf{q}} | \ll q_{c}$
 \begin{equation}
 G ( {\bf{k}}^{\alpha} + {\bf{q}} , i \tilde{\omega}_{n} )
 = \frac{1}{ i \tilde{\omega}_{n} -  {\bf{v}}^{\alpha} \cdot {\bf{q}}
 +  sign ( \tilde{\omega}_{n} ) \frac{i}{2 \tau^{\alpha} } }
 \; \; \; ,
 \label{eq:GMatsubaradisbos}
 \end{equation}
where 
 $\frac{1}{\tau^{\alpha}} = 
 \frac{   | {\bf{v}}^{\alpha} | }{ \ell^{\alpha} }$
is the inverse elastic lifetime.
It is easy to see that this result agrees with
the usual perturbative expression for the average self-energy
in lowest order Born approximation,
which is given by\cite{Abrikosov63}
 \begin{equation}
  Im \Sigma (  {\bf{k}} )
  =    \frac{\gamma_{0}}{V} \sum_{\bf{q}} 
 e^{ - \frac{ | {\bf{q}}|_{1} }{q_{c}} }
 Im G_{0} ( {\bf{k}} + {\bf{q}} , - i 0^{+} )
 \; \; \; .
 \label{eq:Imsigmadis2}
 \end{equation}
Because for $q_{c} \ll k_{F}$ only wave-vectors  with $ | {\bf{q}}| 
 \ll k_{F}$ 
contribute, it is allowed  
approximate the Green's function on the right-hand side of
Eq.(\ref{eq:Imsigmadis2}) by its linearized form.
Then we obtain
 $\frac{1}{ \tau^{\alpha}} 
 =
 2 Im \Sigma ( {\bf{k}}^{\alpha} ) 
 =
 \left( \frac{ q_{c}}{  {\pi}} \right)^{d-1}
 \frac{\gamma_{0}}{ | {\bf{v}}^{\alpha} |}$,
in agreement with the above bosonization result.
In Ref.\cite{Kopietz95May} we have shown that a Debye-Waller
factor which diverges at large distances stronger than
logarithmic completely washes out any singularity in the
momentum distribution $n_{\bf{k}}$.
Hence the average momentum distribution is for any finite disorder
analytic at the Fermi surface. Of course, this is a well known result\cite{Abrikosov63}.
It is also easy to understand 
why in one-dimensional interacting Fermi systems any finite disorder destroys 
the algebraic singularity of $n_{\bf{k}}$,
which is one of the characteristics of a Luttinger liquid\cite{Haldane81}.
Recall that this algebraic singularity 
is due to the logarithmic divergence of
$Q^{\alpha}_{int} ( {\bf{r}} , 0 )$ 
for $r^{\alpha} \rightarrow \infty$.
At sufficiently large distances this weak logarithmic divergence
is negligible compared with the linear divergence
of $Q^{\alpha}_{dis} ( {\bf{r}} , 0 )$.

The case of a time-dependent random potential is more interesting. 
To calculate the average Green's function, we should
specify the dynamic covariance function $C_{q}$.
If we would like to model an underlying interacting
many-body system in thermal equilibrium by a random system, then the form of $C_{q}$ is
completely 
determined by the nature of the interaction. 
In the case of the coupled electron-phonon system at high
temperatures an explicit microscopic calculation
of $C_{q}$ has been given by Girvin and Mahan\cite{Girvin79}.
However, their 
identification of $C_{q}$ with
the parameters of the underlying many-body system 
is based on a perturbative calculation
of the self-energy at high temperatures.
{\it{Our functional bosonization approach allows us 
to relate the covariance function $C_{q}$ of the 
random system in a much more direct and essentially
non-perturbative way to the underlying many-body system.}}
Evidently, the requirement that the average Green's function of the stochastic model
should be identical with the Green's function of the
many-body system without disorder
is equivalent with
$Q^{\alpha}_{dis} ( {\bf{r}} , \tau )
=Q^{\alpha}_{int} ( {\bf{r}} , \tau )$,
where
$Q^{\alpha}_{dis} ( {\bf{r}} , \tau )$ and
$Q^{\alpha}_{int} ( {\bf{r}} , \tau )$ are given 
Eqs.(\ref{eq:QdisDW}) and (\ref{eq:Qalphaint}).
It immediately follows that the connection between the effective stochastic model
and the interacting many-body system is given by the surprisingly simple relation
 \begin{equation}
 C_{q} = -  f_{q}^{RPA}
 \label{eq:CFRPA}
 \; \; \; .
 \end{equation}
For example, to describe longitudinal acoustic phonons
with dispersion $\omega_{\bf{q}}$ that are coupled to the electrons via the
Coulomb potential $f_{\bf{q}}^{cb} = \frac{ 4 \pi e^2}{ {\bf{q}}^2 }$, we 
should choose the covariance function\cite{Kopietz95pho}
  \begin{equation} 
 C_{q} = -
  \frac{ f^{cb}_{\bf{q}} }{ 1 + f^{cb}_{\bf{q}} {\Pi}_{ph} ( q ) } 
  \;  \; , \; \; 
 {\Pi}_{ph} ( q ) = \Pi_{0} ( q ) +  
 \frac{  \gamma  \omega_{\bf{q}}^2}{ \omega_{m}^2 +  \omega_{\bf{q}}^2 } 
 \label{eq:Piphdef}
 \; \; \; ,
 \end{equation}
where ${\gamma}$ measures the strength of the electron-phonon coupling.
We would like to emphasize that in spite of its apparent
simplicity Eq.(\ref{eq:CFRPA}) is a highly non-trivial result, because it 
is based on a controlled summation
of the entire perturbation series of the many-body problem
via bosonization. 
The crucial point is that {\it{bosonization 
produces an exponential resummation of the perturbation series, 
so that the effect of the interactions on the Green's function can be
expressed exclusively in terms of a Debye-Waller factor $Q_{int}^{\alpha} ( {\bf{r}} , \tau )$, 
which can then be directly compared with the Debye-Waller factor
$Q^{\alpha}_{dis} ( {\bf{r}} , \tau )$ 
due to disorder}}.

Of course, the  dynamic random potential
could also be  due to
some non-equilibrium external forces, 
in which case the above identification with an underlying
many-body system is meaningless.
As an example, let us consider
a Gaussian white noise random potential, 
with covariance given by
 $C_{q} =  
 C_{0} e^{ - \frac{ | {\bf{q}} |_{1}}{q_{c}} }$.
Substituting this into Eq.(\ref{eq:QdisDW})
and taking the limit $V, \beta \rightarrow \infty$,
the integrations can be performed analytically, with the result
 \begin{equation}
 Q_{dis}^{\alpha} ( r^{\alpha} \hat{\bf{v}}^{\alpha} , \tau )
 = 
 \frac{ i W \tau }
 { r^{\alpha} + i | {\bf{v}}^{\alpha} | \tau 
 + i sign( \tau ) q_{c}^{-1} }
 \label{eq:Qdisres}
 \; \; \; ,
 \end{equation}
where $W = 
 \frac{C_{0}}{2 \pi } \left( \frac{ q_{c}}{\pi} \right)^{d-1}$.
Because the Debye-Waller factor vanishes at $\tau = 0$, we have
 $G^{\alpha} ( {\bf{r}} , 0 ) = 
 G_{0}^{\alpha} ( {\bf{r}} , 0 )  $, so that
the dynamic white noise random potential 
does not affect the momentum distribution $n_{\bf{k}}$.
It is easy to see that this is an artifact of the white noise limit.
Note, however, that for
$\frac{\tau}{ r^{\alpha} } \rightarrow  \infty$
 \begin{equation}
 G^{\alpha} ( {\bf{r}} , \tau )
 \sim
 -  \delta^{(d-1)} ( {\vec{r}}^{\alpha}_{\bot}  ) \frac{  
 e^{ 
    W / | {\bf{v}}^{\alpha} | }}
 { 2 \pi | {\bf{v}}^{\alpha} | \tau }
 \; \; \; .
 \label{eq:Ginttauinf}
 \end{equation}
The above limit determines 
the density of states at the Fermi energy, which in turn
can be expressed in terms of a renormalized effective mass.
Hence, the random potential enhances
the effective mass by a factor of
 $e^{ W / | {\bf{v}}^{\alpha} | }$.
This is intuitively clear:
the electrons become heavier because 
they have to overcome the resistance of the random potential.
The Fourier transformation of
$G^{\alpha} ( {\bf{r}} , \tau )$ can be calculated exactly.
For $\hat{\bf{v}}^{\alpha} \cdot {\bf{q}} \geq 0$ the result is
 \begin{eqnarray}
 G^{\alpha} ( {\bf{q}} , i \omega )  & = &
 \frac{1}{ W  q_{c} + 
 i \omega - {\bf{v}}^{\alpha} \cdot {\bf{q}}  }
 \nonumber
 \\
 &  & \hspace{-20mm} \times
 \left\{  1 +
 \frac{ W q_{c} e^{ -  {\hat{\bf{v}}}^{\alpha} \cdot {\bf{q}} / q_{c} }}
 { i \omega - {\bf{v}}^{\alpha} \cdot {\bf{q}}   }
 \exp \left[   -   W 
  \frac{ \hat{\bf{v}}^{\alpha} \cdot {\bf{q}} }
 { i \omega - {\bf{v}}^{\alpha} \cdot {\bf{q}}   } \right]
 \right\}
 \label{eq:Gdisqomegares}
 \; .
 \end{eqnarray}
If we now analytically continue this expression to real frequencies
by replacing $ i \omega \rightarrow \omega + i 0^{+}$, we encounter
an essential singularity at $\omega = {\bf{v}}^{\alpha} \cdot {\bf{q}}$.
We suspect that this is an artifact of the
above simple choice of the random potential.

In summary,
we have shown that functional bosonization
can be used to calculate the disorder averaged Green's function
of electrons at high densities that are subject to
a random potential with long-range spatial correlations.
While in the static limit we have recovered the 
usual perturbative result, for time-dependent random potentials
we have obtained a highly non-trivial expression
for the averaged single-particle Green's function.
One of our main results is Eq.(\ref{eq:CFRPA}), which
puts the description of an interacting many-body system
via an effective stochastic model on a solid microscopic basis. 
Although our method  describes  disorder and interactions on equal footing,
the corresponding contributions to the Debye-Waller factors  
are simply additive, so that
interference terms containing
weak localization effects do not appear.
However, it might be possible 
to calculate the single-particle Green's function
via higher-dimensional bosonization beyond the Gaussian approximation,
taking momentum-transfer  between different patches
or the non-linearities in the energy dispersion
approximately into account.
For the density-density correlation function
and the boson representation of the Hamiltonian such a calculation
has been performed in Ref.\cite{Kopietz94b}.
In this case bosonization in $d > 1$ could become a new powerful
approach to disordered interacting Fermi systems in the diffusive regime, which 
can deal simultaneously with disorder and interactions and 
does not have the disadvantages of the replica trick.
Note that the diffusive regime does not exist in $d=1$, so that
the physics of weak localization can only be 
discussed within higher-dimensional bosonization.
Work in this direction is in progress.

The inclusion of transverse gauge fields is straightforward within
our functional bosonization approach.
The results of the present work can be combined 
with the method described in Ref.\cite{Kopietz95March} to treat the problem of a random transverse
gauge field with long-range spatial correlations in arbitrary dimensions.
This problem
is of experimental relevance in connection with
the quantum Hall effect\cite{Ludwig94}.
It should be mentioned, however, that
the applicability of higher-dimensional bosonization 
to transverse gauge fields is controversial\cite{Altshuler94}.
We hope to address this issue in future publications.

I would like to thank G. Castilla for carefully reading the
manuscript and suggesting many useful modifications.
I am also grateful to
Kurt Sch\H{o}nhammer and Volker Meden for discussions.

\end{document}